\documentclass[twoside,fleqn]{article}

\usepackage{espcrc2}

\input epsf

\newcommand{\AmS}{{\protect\the\textfont2
  A\kern-.1667em\lower.5ex\hbox{M}\kern-.125emS}}

\title{Deflation of Eigenvalues for Iterative Methods in Lattice QCD}

\author{Dean Darnell\address{Department of Physics, Baylor University, Waco, TX 76798-7316},
Ronald B. Morgan\address{Department of Mathematics, Baylor University, Waco, TX 76798-7328}
and Walter Wilcox$^a$}

\begin{document}

\begin{abstract}
Work on generalizing the deflated, restarted GMRES algorithm,
useful in lattice studies using stochastic noise methods, is reported. 
We first show how the multi-mass extension of deflated GMRES 
can be implemented. We then give a deflated GMRES method that 
can be used on multiple right-hand sides of $Ax=b$ in an efficient manner. 
We also discuss and give numerical results on the possibilty of 
combining deflated GMRES for the first right hand side with a 
deflated BiCGStab algorithm for the subsequent right hand sides. 

\vspace{1pc}

\end{abstract}


\maketitle

\section{INTRODUCTION}

Removal of the smallest eigenvalues of a matrix can significantly improve 
the performance of iterative methods. This is important for lattice QCD where 
stochastic methods for disconnected diagrams and new matter formulations such as 
overlap fermions are computationally intensive.  In addition, there is often a need 
for these systems to be shifted to take account of multiple quark masses or other 
parameters.  We think that it is important for any iterative method we develop 
to be applicable to general non-Hermitian systems, for this is often the case for lattice systems.

GMRES~\cite{1} is an iterative method that can be used in lattice studies with 
stochastic noise methods.  Convergence can be improved by deflating eigenvalues.  Two years
ago it was shown that deflating eigenvalues~\cite{new} can be especially helpful for lattice
problems with multiple right-hand sides~\cite{2}.  Here we report on generalizations of these
techniques.  First we show how deflated restarted GMRES can accomodate multiple shifts.  We give
numerical results for a Wilson matrix with very small eigenvalues.  We then consider the case of
combining multiple shifts with multiple right-hand sides.  Finally we also discuss and give
numerical results on the possibility of combining deflated GMRES for the first system with a
BiCGStab algorithm which uses those deflated eigenvalues for the subsequent solutions.

\section{DEFLATION FOR MULTIPLY SHIFTED SYSTEMS}

The shifted systems are expressed by
\begin{eqnarray}
(A - s(i)I)x^i = b, 
\end{eqnarray}
where the s(i) are the numerical shifts. The method proposed in \cite{3} is
based upon keeping the residual vectors for the shifted systems parallel 
to one another. That is, one uses only a single Krylov subspace to solve all 
the systems. This is possible for GMRES only by forcing the residuals to be 
parallel to one another after a restart and thus the minimum residual condition 
is applied to only the unshifted (most difficult) system. We find that this also 
results in an almost optimal solution for the shifted systems.

\begin{figure}
\begin{center}
\leavevmode
\epsfxsize=2.9 in
\epsfbox{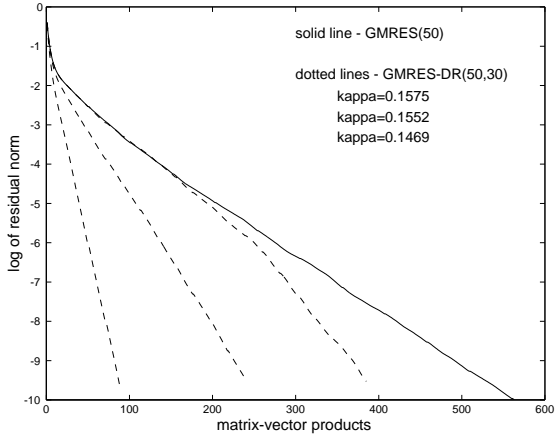}
\vskip-.1cm
\caption{ Comparison of the residual norms for standard 
(unshifted) GMRES(50) (solid magnenta) with shifted GMRES-DR(50,30).  This is 
for a $12^3\times 24$ Wilson-Dirac matrix at $\kappa=0.1575$ (unshifted), 0.1552, and
0.1469.\label{figure1}}
\end{center}
\vskip-.5cm
\end{figure}
Our method for deflating GMRES (called GMRES-DR~\cite{2}) simulataneously solves linear equations
and computes eigenvalues. This is possible because it has a Krylov subspace for solving the linear
equations as well as Krylov subspaces designed for solving the eigenvalue equation. These
approximate eigenvectors are refined 
and improved as the GMRES-DR cycles proceed.  We have determined that
GMRES-DR can be efficiently applied to multiply shifted systems because of its Krylov properties.  

Fig.1 shows the result of a shifted run of GMRES-DR(50,30), i.e., the subspace is 
50-dimensional and includes 30 approximate eigenvectors.  It is compared to the residual norm for
regular GMRES(50). One can see the deflation start to 
take place around 200 matrix-vector products (the approximate eigenvectors are accurate 
enough by then). 

\section{MULTIPLE SHIFTS AND ADDITIONAL RIGHT-HAND SIDES}

Typically, many right-hand sides are used to do stochastic estimates of matrix elements.  
It would be beneficial if the deflated eigenvalues and eigenvectors 
calculated for the first right-hand side could be reused. 
Then deflation could begin from the start instead of having to wait for the 
required eigenvectors to be recalculated, with more dramatic results than in Fig.1. A
method to accomplish this called GMRES-Proj was given in~\cite{2}.  GMRES-Proj alternates cycles of
GMRES with projections over the approximate eigenvectors.  Next we consider using this for the case
of multiple shifts for each of the multiple right-hand sides.

GMRES-DR can deflate eigenvalues and handle multiple shifts at the same time for the first
right-hand side (the right-hand sides for each shifted system can be kept parallel).  However, for
the second and subsequent right-hand sides there is no way to deflate
eigenvalues by projecting over the previously computed approximate eigenvectors and still keep the
right-hand sides for the different shifted systems parallel.  If the eigenvectors were exact, this
would be possible.  So one approach is to use only fairly accurate approximate eigenvectors and do a
correcting iteration at the end for each shifted system separately.  

We now propose a better fix for this problem which does not require accurate eigenvectors. 
GMRES-DR produces approximate eigenvectors that are represented in the form of an Arnoldi-like
recurrence relation:
\begin{eqnarray}
 AV_k = V_{k+1}\bar H_k, 
\end{eqnarray}
where the columns of $V_k$ span the space of approximate eigenvectors, $V_{k+1}$ has one extra
column added to $V_k$, and $\bar H_k$ is a $(k+1)$ by $k$ matrix.  Using this, we can perform the
projection over the approximate eigenvectors so that the right-hand sides for the different shifted
systems are all multiples of each other, except for error in the direction of $v_{k+1}$, the last
column of $V_{k+1}$.  This error can be easily corrected at the end, if we are willing to solve one
additional right-hand side, namely $v_{k+1}$.  

There is also potential to develop a deflated version of BiCGStab for multiply 
shifted systems using this idea. This will be reported later, but for now we will look at deflating
BiCGStab for multiple right-hand sides in the non-shifted case.

\section{DEFLATED BICGSTAB}	

The deflation on the additional right-hand sides can employ other methods 
after the initial eigenvalue generation by GMRES-DR. We have looked at 
BiCGStab, a non restarted method popular in the lattice community, in 
this context. We again combine a projection over approximate 
eigenvectors with the iterative method. However, the projection can 
only be performed once before the BiCGStab iteration begins. Therefore 
this projection must reduce the critical eigencomponents to a low enough 
level that no further reduction is needed during the BiCGStab iteration.
The minimal residual projection is the best possible one in the sense that 
it minimizes the norm of the residual vector.  However, it can be shown that even 
if one knows the exact eigenvectors, the corresponding component in the residual vector will not
generally be zeroed out (unless the matrix is Hermitian). To deal with this problem, we will give a
projection that is better for this task. It uses both left and right approximate eigenvectors.

\vspace{.2cm}
Left-right Projection for deflating eigenvectors:

\vspace{.2cm}
1. Let the current approximate solution be $x_0$ and the current system of 
equations be $A(x-x_0) = r_0$. Let $V$ be orthonormal with columns spanning the 
subspace of approximate right eigenvectors. Let $U$ similarly be orthonormal 
with columns from approximate left eigenvectors. 

\vspace{.2cm}
2. Solve $U^{\dagger}AVd = c$, where $c = U^{\dagger}r_0$.
\vspace{.2cm}

3. The new approximate solution is $ x_k = x_0 + Vd$.

\vspace{.2cm}
This projection can reduce the small eigenvector components better 
than minimum residual and in the case of exact left and right eigenvectors, 
the components can be zeroed out. Of course, it is necessary to compute left 
eigenvectors. However, for many lattice problems, 
including the Wilson-Dirac matrix, there is a simple relationship between 
the left and right eigenvectors.

\begin{figure}
\begin{center}
\leavevmode
\epsfxsize=2.9 in
\epsfbox{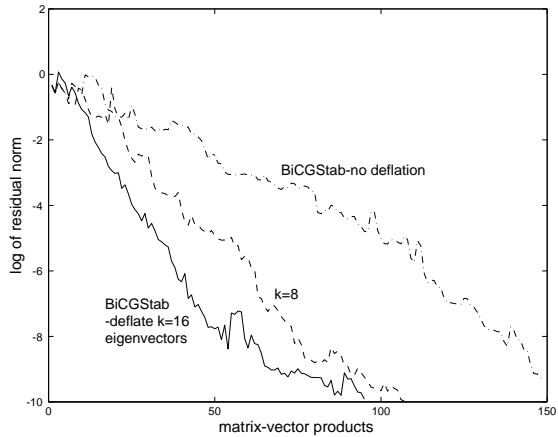}
\vskip-.1cm
\caption{Comparison of standard nondeflated BiCGStab with deflated versions 
using either 8 or 16 eigenvectors from the initial GMRES-DR run.\label{figure2}}
\end{center}
\vskip-.5cm
\end{figure}

Fig.2 above shows a run on a smaller $4^4$ lattice (using MATLAB) of the 
deflated BiCGStab algorithm using eigenvectors calculated from an initial 
GMRES-DR solution of the first right-hand side. The deflated solutions take 
a smaller number of iterations than the nondeflated implementation. 
Notice that deflation does not care about the nature of the right-hand 
sides used, so these may be noise vectors or units vectors, as one desires. 
This applies to all techniques used.

\section{ACKNOWLEDGEMENTS}

This material is based upon work supported by the National Science Foundation under Grants No.
0070836 (Theoretical Physics) and 0310573 (Computational Mathematics) and NCSA and
utilized the SGI Origin 2000 System at the University of Illinois. RM acknowledges the Baylor
University Summer Sabbatical Program.

\end{document}